\documentclass[aps,twocolumn,superscriptaddress]{revtex4}
\usepackage{times}
\usepackage{color}
\usepackage{float}
\usepackage{amssymb}
\usepackage{epsfig}
\usepackage{mathrsfs}
\usepackage{amsmath}
\usepackage{mathrsfs}
\usepackage{graphicx}

\DeclareMathAlphabet{\mathpzc}{OT1}{pzc}{m}{it}

\begin{document}

\title{Novel transition and Bellerophon state in coupled Stuart-Landau oscillators }

\author{Jiameng Zhang}
\affiliation{Department of Physics, East China Normal University, Shanghai, 200241, China}

\author{Xue Li}
\affiliation{Department of Physics, East China Normal University, Shanghai, 200241, China}

\author{Yong Zou}
\affiliation{Department of Physics, East China Normal University, Shanghai, 200241, China}

\author{Shuguang Guan}\thanks{Corresponding author: guanshuguang@hotmail.com}
\affiliation{Department of Physics, East China Normal University, Shanghai, 200241, China}

\begin{abstract}
We study synchronization in a system of Stuart-Landau oscillators with frequency-weighted coupling. For three typical unimodal frequency distributions, namely, the Lorentzian, the triangle, and the uniform, we found that the first-order transition occurs when the frequency distribution is relatively compact,  while  the synchronization transition is continuous when it is relatively wide. In both cases, there is a regime of Bellerophon state between the incoherent state and the synchronized state. Remarkably, we revealed novel transition behavior for such coupled oscillators with amplitudes, i.e., the regime of Bellerophon state actually contains two stages. In the first stage, the oscillators achieve chaotic phase synchronization; while in the second stage, oscillators form periodical phase synchronization.
Our results suggest that Bellerophon state also exists in coupled oscillators with amplitude dynamics.
\end{abstract}

\maketitle

\section{Introduction}

Synchronization is a type of macroscopic order self-organized in dynamical systems consisting of two or more interacting units. Such phenomena turn out to be ubiquitous in nature, such as in physics, chemistry, life and biology, engineering, and social science \cite{Pikovsky2003}. For example, typical synchronization includes
the flashing of fireflies \cite{Buck1988},
the circadian rhythms of plants and animals \cite{Winfree1967},
neurons in human brain \cite{Buzski2004},
power grid \cite{Rohden2012},
the Josephson junction arrays \cite{Wiesenfeld1996},
and the crowd synchrony on the Millennium Bridge \cite{Eckhardt2007}, etc.
Since synchronization is the dynamical basis for  cooperative functioning in a wealth of systems, it has been extensively investigated for the past several decades, both theoretically and experimentally.

One classical, and also the most successful, model for theoretical study on synchronization is the Kuramoto model \cite{Kuramoto1984,Strogatz2000}:
\begin{eqnarray}\label{eq:KM1}
 		\dot{\theta}_j  = \omega_j + \frac{K}{N}\sum_{n=1}^N \sin(\theta_n -\theta_j)  \quad j=1,2,\cdots,N,
\end{eqnarray}
where $\theta_j$ and  $\omega_j$ are the phase and the natural frequency of the $j$th phase oscillator, and the dot denotes the time derivative. The second term of the right hand side is the coupling among oscillators. Basically, this model
describes the synchronization transition among a large number of phase oscillators via mean-field coupling. On one hand, it is simple enough to apply analytical treatment. On the other hand, it can capture the most fundamental dynamics in coupled oscillators. Due to these reasons, the Kuramoto model and its many variants have been extensively investigated for over forty years, which has greatly enhanced our understandings about the collective behaviors of coupled systems. For a latest and comprehensive review, please refer to Ref. \cite{Rodrigues2016}.

One central issue in the research of synchronization in coupled oscillators is the formed coherent state, which emerges autonomously due to the nonlinear interaction among oscillators. So far, studies have revealed various coherent states in Kuramoto-like models. Essentially, they can be classified into two types: the stationary and the non-stationary.
In the continuum limit, i.e., the number of oscillators $N\to \infty$, a density function $\rho(\theta,\omega,t)$ can be introduced, such that $\rho(\theta,\omega,t) \ d\theta$ accounts for the fraction of oscillators of natural frequency $\omega$ whose phases are between $\theta$ and  $\theta+d\theta$ at time $t$. $\rho$ satisfies the normalization condition
\begin{equation}
\int_0^{2\pi}\rho(\theta,\omega,t)d\theta=1
\end{equation}
for all $\omega$ and all $t$, and its evolution is governed by the continuity equation
\begin{equation}\label{eq:continuity}
\frac{\partial\rho}{\partial t}+\frac{\partial (\rho\upsilon)}{\partial \theta}=0.
\end{equation}
For a coherent state, if the corresponding density does not change with time, it is defined as stationary state; on the contrary, if the density of a coherent state varies with time, it is non-stationary. Typical stationary states include the (partially) synchronized state (or the coherent state) \cite{Kuramoto1984,Strogatz2000}, the $\pi$-state \cite{Hong2011}, and the travelling wave state (observed in an appropriate rotating frame) \cite{Iatsenko2013}, while the standing wave state belongs to the non-stationary state \cite{Crawford1994}.

Recently, investigations have identified two novel coherent states in coupled phase oscillators. One is the Chimera state (CS) , which consists of both coherent and incoherent dynamics though in coupled identical oscillators with symmetric coupling \cite{Kuramoto2002,Omelchenko2013,Abrams2008}. 
The other is the Bellerophon state (BS) , which is a quantized and non-stationary coherent phase, occuring in  globally coupled nonidentical oscillators with widely different frequencies \cite{Boccaletti2016,Bi2016,Zhou2016,Bi2017,Qiu2016,Qiu2018}.
In such state,
oscillators form quantized coherent clusters, and in each coherent cluster the oscillators' instantaneous frequencies are not locked, but instead their average frequencies are locked to form a staircase-like structure.

The BS was first observed in generalized Kuramoto models: either in the frequency-weighted Kuramoto model \cite{Bi2016,Zhou2016,Bi2017} or in the Kuramoto model with conformists and contrarians \cite{Qiu2016,Qiu2018}. Later it was revealed that it could also occur in classical Kuramoto model with bimodal frequency distribution \cite{Li2019}. Therefore, the BS is in fact a generic organization of globally coupled nonidentical phase oscillators occurring at intermediate values of the coupling strength, not limited to specific dynamical model nor to special arrangements in the frequency distributions. We noticed that so far the BS has been only observed in coupled phase oscillators, where the dynamics of each oscillator is greatly simplified to be described only by a phase variable. However, in many real dynamical systems, the amplitude usually plays a crucial role that cannot be ignored. Then one natural question is: could the BS occur in coupled oscillators with amplitude dynamics?

Motivated by this idea, in this work we investigate a model of coupled Stuart-Landau (SL) oscillators, which are typical limit cycles with amplitude dynamics.
For a variety of frequency distributions, such as the Lorentzian, the triangle, and the uniform,  we find that BS indeed occurs in this model. By extensive numerical simulations, the formed BSs are fully characterized. Moreover, different synchronization paths, including both the first- and the second-order transitions, have been characterized. Remarkably, we reveal that actually there are two stages within the regime of BS: one is chaotic phase synchronization and the other is periodical phase synchronization.
This work demonstrates that BS might be more generic in the sense that they could also form in coupled oscillators with amplitude dynamics. Hopefully, the present results will stimulate physicists to further seek higher order coherent states in other numerous and diverse conditions and settings.

\section{The dynamical model}

In this work, we study a dynamical model
of globally coupled SL oscillators with frequency-weighted coupling \cite{Bi2016,Zhou2016}, i.e.,
\begin{equation}\label{eq:model}
\dot{z}_j(t)=(a+iw_j-|z_j|^{2})z_j(t)+
\frac{K|\omega_j|}{N}\sum_{n=1}^{N}[z_n(t)-z_j(t)].
\end{equation}  ¡¡
Here $j=1,2,\cdots,N$ denotes the index of oscillators. $z_j(t)=x_j(t)+iy_j(t)$
is the complex amplitude of the $j$th oscillator at time $t$, and the dot represents the time derivative. $a$ is a control parameter for individual SL oscillator, i.e., the dynamics settles on a limit cycle if $a>0$, and on a fixed point if $a<0$.
$\omega_j$ is the natural frequency
of the $j$th oscillator, and $K$ is the coupling strength. Compared with the phase oscillator in classical Kuramoto model, the dynamics of an individual SL oscillator is two-dimensional which has both amplitude and phase.

In this work, we consider typical unimodal frequency distributions, including the Lorentzian, the triangle, and the uniform. Their analytical forms $g(\omega)$ are given as follows.
\begin{enumerate}
\item {Lorentzian distribution.}
\begin{equation}\label{eq:Lorentzian}
g(\omega) = \frac{\Delta}{\pi (\omega ^2 + \Delta ^2)}.
\end{equation}
\item {Triangle distribution.}
\begin{equation}\label{eq:triangle}
g(\omega) =(\pi\Delta-|\omega|)/(\pi\Delta)^2,\quad |\omega|<\pi\Delta, \mbox{0 otherwise}.
\end{equation}
\item {Uniform distribution.}
\begin{equation}\label{eq:uniform}
g(\omega) = 1/(2\pi\Delta),\quad |\omega|<\pi\Delta, \mbox{0 otherwise}.
\end{equation}
\end{enumerate}
In all three distributions, $\Delta$ is a parameter which controls the width of the distribution.

To characterize the collective behaviors of the coupled SL oscillators, two order parameters can be defined as:
\begin{equation}\label{eq:op1}
R_ze^{i\psi}=\sum_{j=1}^{N}z_j/N,
\end{equation}
and
\begin{equation}\label{eq:op2}
R_{\theta}e^{i\phi}=\sum_{j=1}^{N}e^{i\theta_j}/N.
\end{equation}
Here, $\theta_j$ represents the phase of the $j$th oscillator.
Order parameter $R_z$  ($0\le R_z \le1$ due to $a=1$ in this study) characterizes the coherence of the complete dynamics, including both amplitude and phase.  Order parameter $R_{\theta}$ ($0\le R_\theta \le 1$) only characterizes the phase coherence of the system, which does not involve any information of amplitude.

Without losing generality, we set $a=1$, $N=2000$, and only consider the situation of global coupling.
Throughout this paper, numerical integration is carried out by the fourth-order Runge-Kutta method with time step 0.005. The initial phases of the limit cycles are random, i.e., oscillators are uniformly distributed on the unit circle in complex plane at the beginning.
In our study, both the forward and backward transitions are numerically investigated in an adiabatic way with $\Delta K=0.01$.
To compute the order parameters and other statistical measures, Eq. (\ref{eq:model}) is first integrated for a transient period of $t=10000$. Then the quantities are calculated based on the following time window of length $t=200$. Such scheme is adopted throughout this paper.

\begin{figure}[htbp]
\begin{center}
\includegraphics[width=0.5\textwidth]{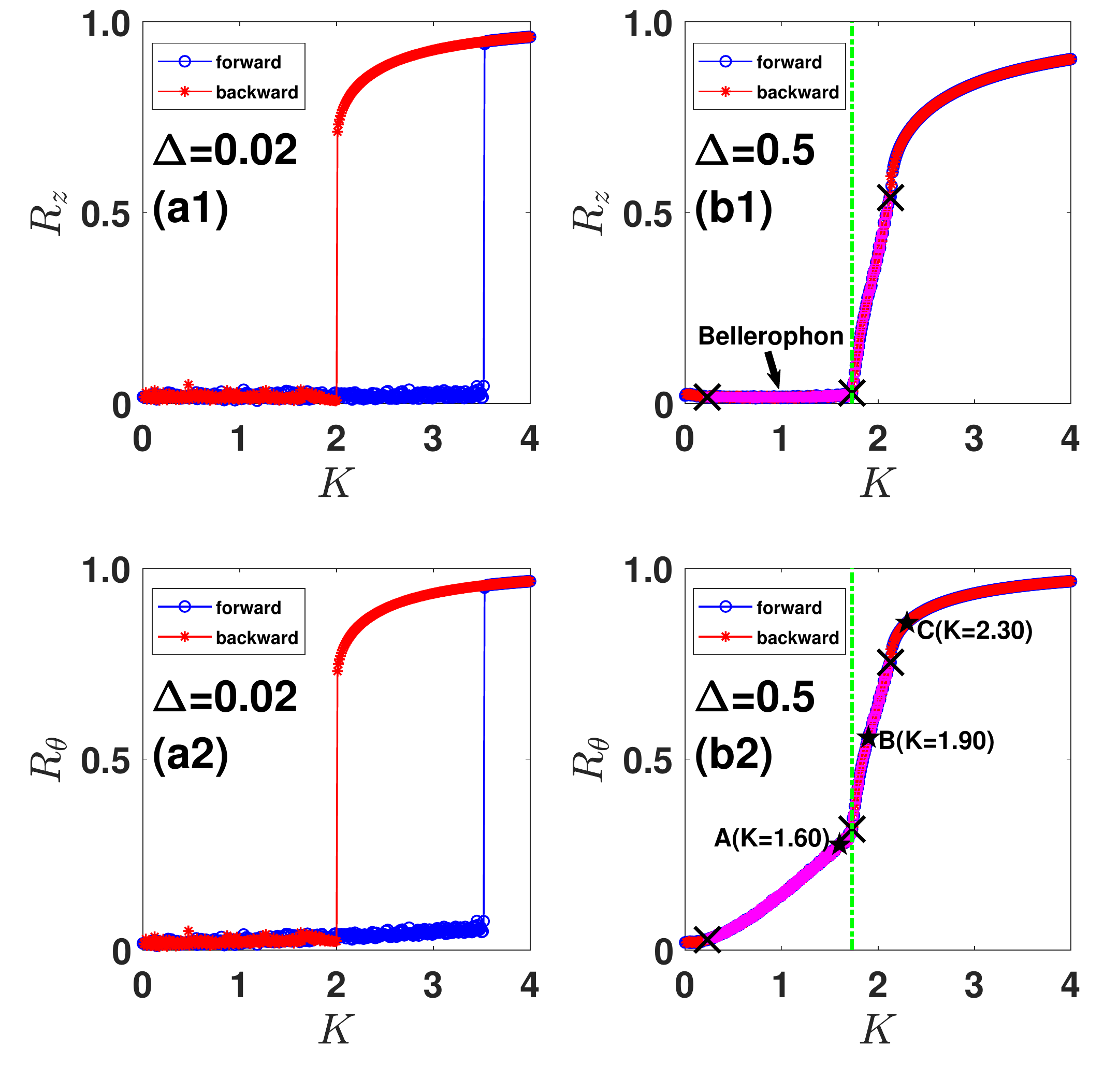}
\caption{(Color online).
Typical synchronization transition paths in coupled SL oscillators (Eq. \ref{eq:model}) with Lorentzian frequency distribution. Both order parameters $R_z$ and $R_\theta$ are plotted vs the coupling strength $K$.
(a) The first-order transition under $\Delta=0.02$. The critical points for the forward and backward transitions are $K$=3.52 and $K=$1.99, respectively.
(b) The second-order transition under $\Delta=0.5$. Three bifurcation points (marked by the crosses) are $K$=0.23, 1.73, and 2.13, respectively. The regime of BS are denoted by the magenta color and the green line marks the transition point within BS regime.
}\label{fig1}
\end{center}
\end{figure}

\begin{figure}[htbp]
\begin{center}
\includegraphics[width=0.5\textwidth]{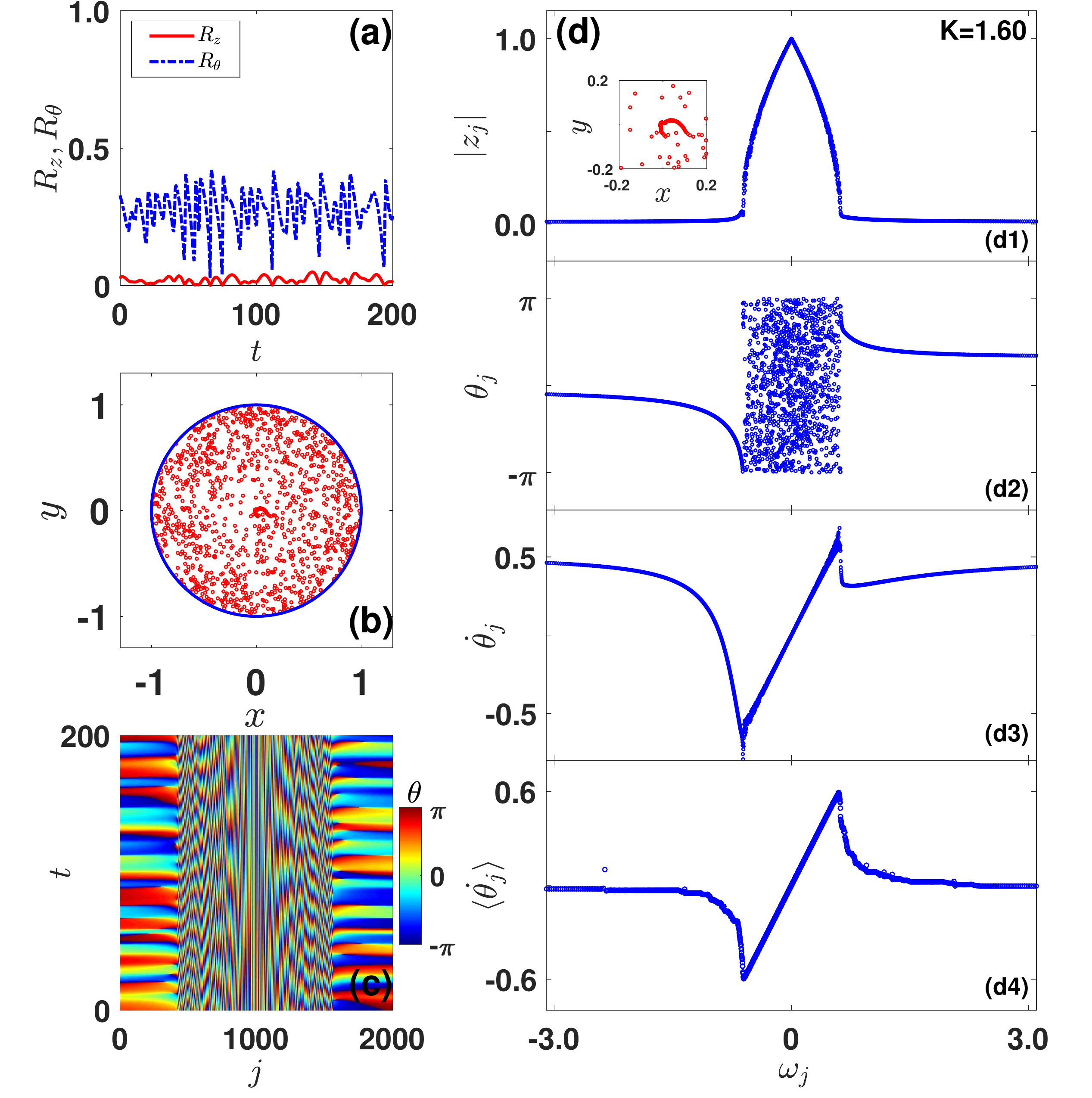}
\caption{(Color online).
The BS corresponding to point A ($K=1.60$) in Fig. \ref{fig1}(b2).
(a) Order parameters $R_z(t)$ and $R_\theta(t)$.
(b) The snapshot of oscillators in the complex plane.
(c) The spatiotemporal pattern of the phases of oscillators, i.e., $\theta(j,t)$, where $j$ denotes the oscillator index.
(d) Snapshots of the instantaneous magnitude of amplitude $z_j$ (d1), the instantaneous phase $\theta_j$ (d2), the instantaneous speed  $\dot{\theta}_j $ (d3), and
the average speed $\langle \dot{\theta}_j \rangle$ (d4) vs the natural frequencies $\omega_j$ of the oscillators.
The inset in (d1) is the enlargement of (b) in a small region around the origin.
}\label{fig2}
\end{center}
\end{figure}

\begin{figure}[htbp]
\begin{center}
\includegraphics[width=0.5\textwidth]{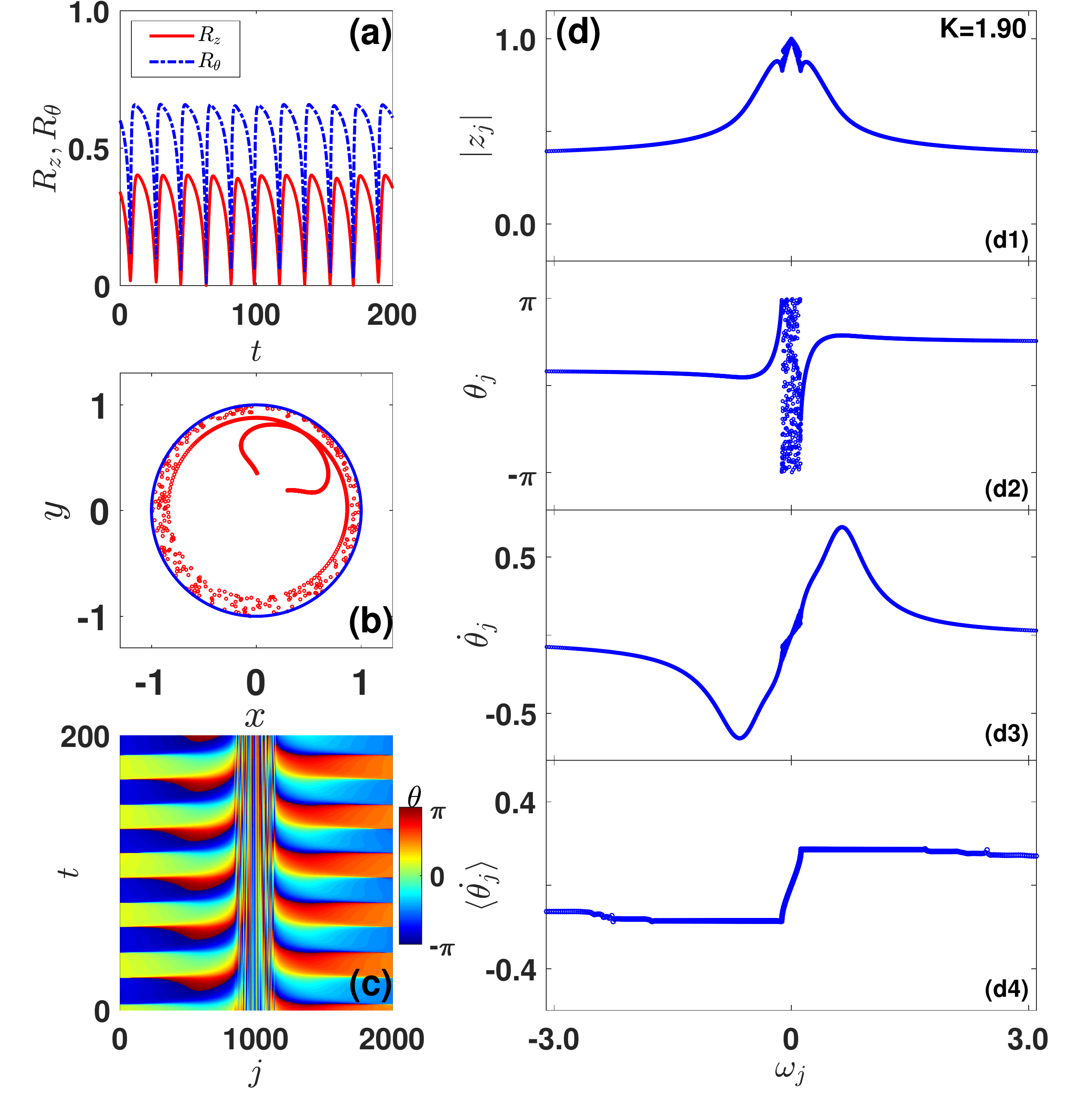}
\caption{(Color online).
The BS corresponding to point B ($K=1.90$) in Fig. \ref{fig1}(b2). The figure caption is the same as in Fig. \ref{fig2}.
}\label{fig3}
\end{center}
\end{figure}

\begin{figure}[htbp]
\begin{center}
\includegraphics[width=0.5\textwidth]{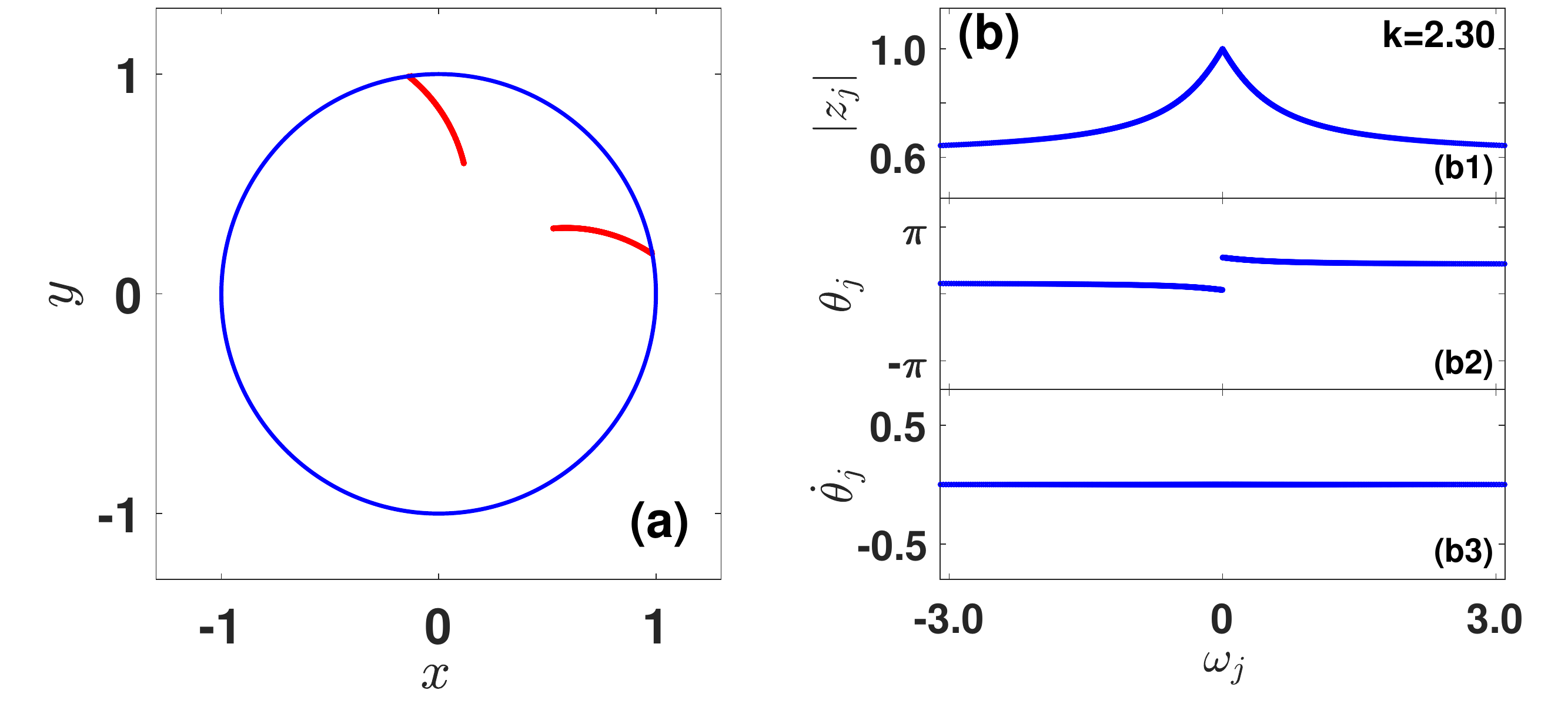}
\caption{(Color online).
The SS corresponding to point C ($K=2.30$) in Fig. \ref{fig1}(b2). The figure caption is the same as in Fig. \ref{fig2}.
}\label{fig4}
\end{center}
\end{figure}

\begin{figure}[htbp]
\begin{center}
\includegraphics[width=0.5\textwidth]{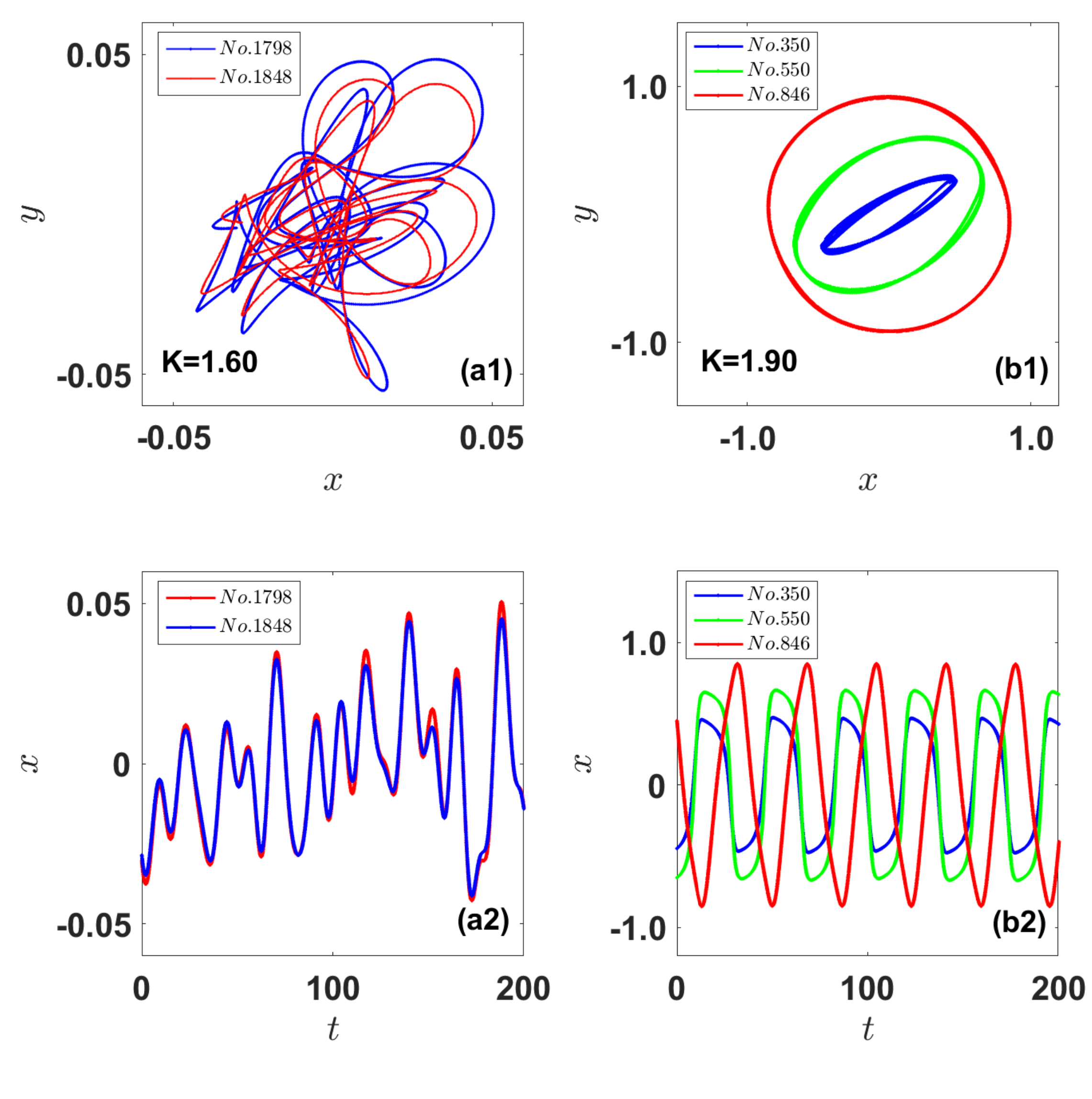}
\caption{(Color online).
The orbits in the complex plane (upper panels) and the time series of $x$ variables of representative oscillators, corresponding to Fig.~\ref{fig2} and Fig.~\ref{fig3}, respectively.
}\label{fig5}
\end{center}
\end{figure}

\section{Results}

In this work, we carry out extensive numerical simulations to investigate the synchronization transitions in the system. Special attention has been paid to the paths towards synchronization and the formed coherent states. In the following, we report the main results in detail.

We first investigate the dynamical model (Eq. \ref{eq:model}) with the Lorentzian frequency distribution, i.e., Eq. (\ref{eq:Lorentzian}). Figure ~\ref{fig1} characterizes the synchronization transitions in this case, where both order parameters $R_z$ and $R_\theta$  are plotted with respect to the coupling strength $K$. It is found that there are three types of attractors in the system with Lorentzian distribution: the incoherent state (IS), the Bellerophon state (BS), and the synchronized state (SS).
When the coupling strength is small, the system evolves into the IS, where no coherent cluster of oscillators is formed.
On the other hand, when the coupling strength is large enough, the oscillators will form SS, where their instantaneous frequencies are locked.
In the intermediated regime of the coupling strength, the system can achieve another higher order coherence, i.e., the BS, where the instantaneous frequencies of oscillators in the cluster are not locked, but their averaged frequencies are locked \cite{Bi2016,Boccaletti2016}.

Depending on parameter $\Delta$ in the Lorentzian distribution, which characterizes the peak width at half height, the system exhibits two scenarios of bifurcations toward synchronization. In the first case, where $\Delta=0.02$, the system undergoes typical first-order transition from the IS to the SS with a hysteresis loop (IS$\rightarrow$SS);
while in the second case, where $\Delta=0.5$, the system bifurcates continuously from the IS to SS via BS (IS$\rightarrow$BS$\rightarrow$SS).

Interestingly, in the regime of BS shown in Fig.~\ref{fig1}(b), we find another transition point (marked by the green line). Comparing Fig.~\ref{fig1}(b1) with (b2), it is seen that the order parameter $R_z$ and $R_\theta$ exhibit different behaviors. For $R_z$, the forward bifurcation point is at $K=1.73$, while for  $R_\theta$, the first forward transition occurs much earlier at $K=0.23$. On the other hand, both $R_z$ and $R_\theta$ show a transition point at $K=1.73$.
So, what happens in the regime of BS, i.e., $K\in[0.23,2.13]$?
A careful examination reveals that in such a regime, the system essentially goes into BS, but there are two slightly different BS!
To highlight this point, we specially choose three representative points at $K=1.60$ (A), $1.90$ (B), and $2.30$ (C) (marked by the stars in Fig.~\ref{fig1}(b2), respectively.
Fig.~\ref{fig2} characterizes the state at point A in Fig.~\ref{fig1}(b2). As seen in (a), the oscillators at this point have formed certain coherent behavior.
However, such coherence is only achieved in the phases, not the amplitudes ((b) and (c)). From (c) and (d), we can see that the coherent cluster contains oscillators with relatively larger natural frequencies, while from (b) and the inset in (d1) we find that the coherent cluster is located in a quite small region around the origin in the complex plane, i.e., the coherent oscillators have very small amplitudes. This directly leads to that $R_z \approx 0$ while $R_\theta$ is significantly greater than 0 as shown in (a). Statistically, except for the small coherent cluster where coherent oscillators have very small amplitude, most oscillators distribute almost randomly inside the unit circle (b). This explains that during the interval $K \in [0.23, 1.73]$, $R_z \approx 0$ while $R_\theta>0$ (Fig. \ref{fig1}(b)).

As the coupling strength further increases, the amplitudes of oscillators gradually become coherent. Fig.~\ref{fig3} shows the state at point B in Fig.~\ref{fig1}(b2). As seen in (b) and (d), there are two coherent clusters in which both phases and amplitudes of oscillators are significantly concentrated. These two coherent clusters rotate in opposite directions, but inside each coherent cluster each oscillator has its own, time-dependent speed (instantaneous frequency). For this reason, both order parameter $R_z$ and $R_\theta$ are oscillating, and significantly greater than 0 on average (a).

As the coupling strength finally increases to exceed the transition point at $K=2.13$, it is found that oscillation death (OD) occurs in this system. Fig.~\ref{fig4} illustrates the synchronized state at point C in Fig.~\ref{fig1}(b2), where all oscillators split to form two coherent clusters consisting of fixed points.

In order to reveal the mechanism underlying the novel transition within the BS regime, we turn to study the microscopic dynamics of individual oscillator. Fig.~\ref{fig5}(a) shows the time evolution of two arbitrary coherent oscillators corresponding to Fig.~\ref{fig2}. We find that the orbits of oscillators evolve in a chaotic way (a1), and the two coherent oscillators actually achieve chaotic phase synchronization, i.e., their phases are locked, but amplitudes are uncorrelated (a2). On the contrary,  in Fig.~\ref{fig5}(b), which corresponds to Fig.~\ref{fig3},
the dynamics of oscillators in this stage become smeared limited cycles with different amplitudes. Those are almost periodic oscillations, which have achieved evident phase synchronization.

Based on the above analysis, we now understand the remarkable transition within the BS regime in  Fig.~\ref{fig1}(b). Actually, the BS regime for the case of $\Delta=0.5$ includes two qualitatively different stages.
In the first stage, the individual oscillator behaves chaotically. Their phases have achieved coherence, but the amplitudes do not. This is a typical situation of chaotic phase synchronization. Then with further increasing of coupling strength, the dynamics of oscillators become periodic and phase synchronization can be achieved.

\begin{figure}[htbp]
\begin{center}
\includegraphics[width=0.5\textwidth]{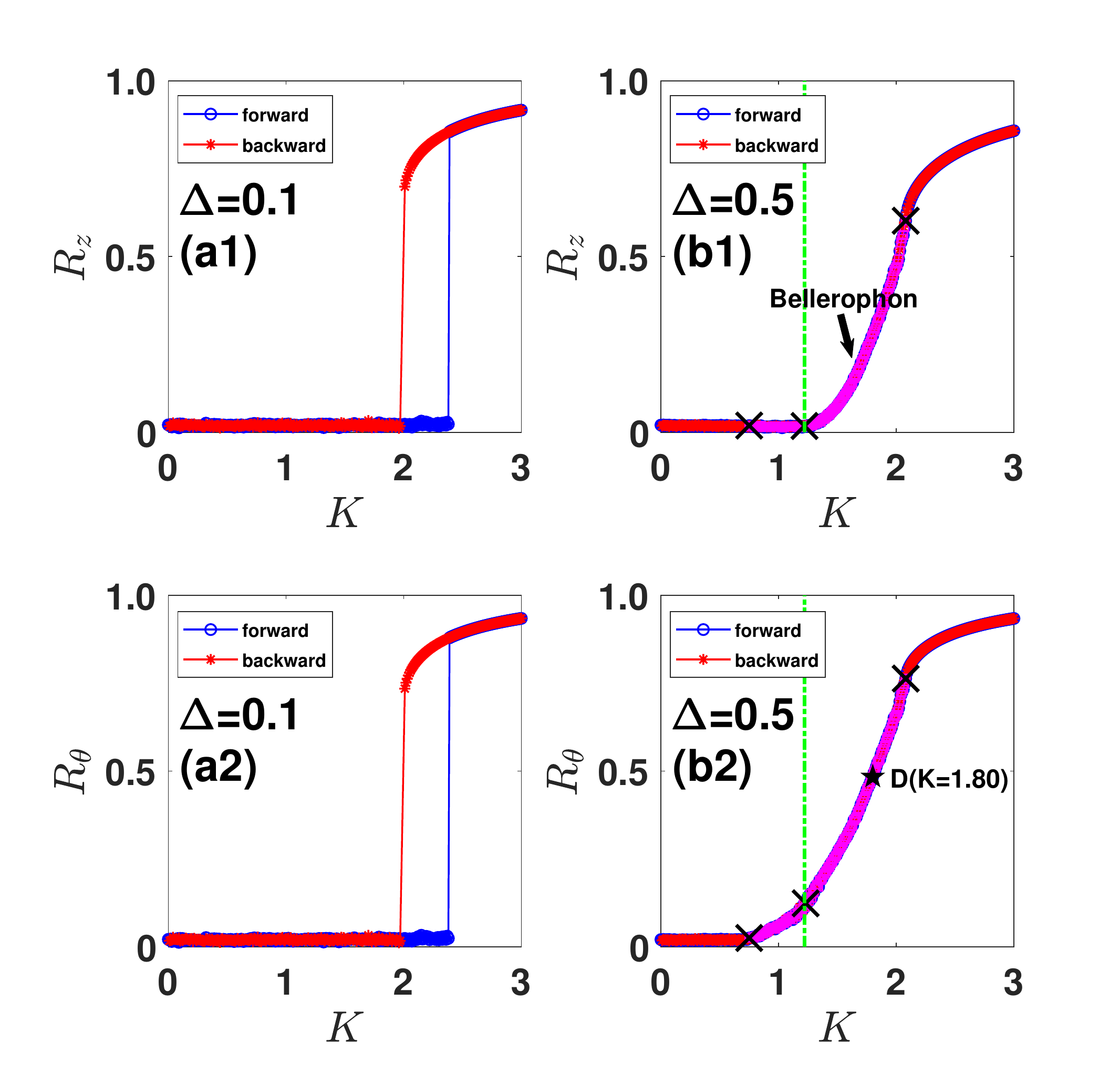}
\caption{(Color online).
Typical synchronization transition paths in coupled SL oscillators (Eq. \ref{eq:model}) with the triangle frequency distribution. For $\Delta=0.1$, the two critical points of forward and backward transitions are $K=2.32$ and $K=2.00$, respectively.
For $\Delta=0.5$, there are three transition points $K=0.74, 1.22$ and $2.08$ (marked by the crosses), respectively.
The regime of BS is denoted by the magenta color, and the green line marks the transition point within BS regime.
}\label{fig6}
\end{center}
\end{figure}

\begin{figure}[htbp]
\begin{center}
\includegraphics[width=0.5\textwidth]{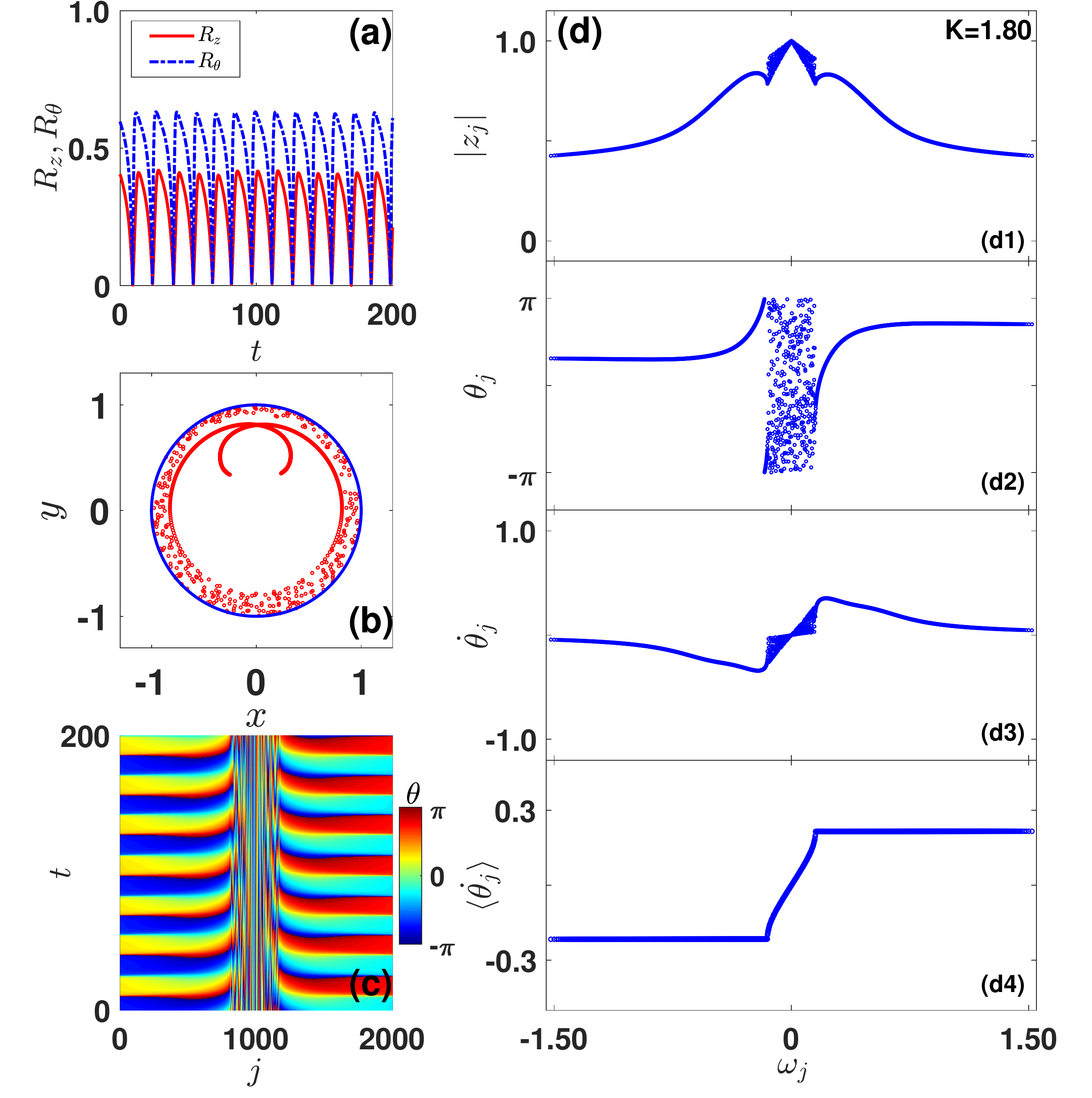}
\caption{(Color online).
The BS corresponding to point D ($K=1.80$) in Fig. \ref{fig6}(b2). The figure caption is the same as in Fig. \ref{fig2}.
}\label{fig7}
\end{center}
\end{figure}

Besides the Lorentzian frequency distribution, we have also studied two other  unimodal distributions, i.e., the triangle distribution and the uniform distribution. In both cases, we observe qualitatively similar results.
Fig. \ref{fig6} shows the bifurcation paths toward synchronization with the triangle frequency distribution.
When parameter $\Delta$ is small, for example, $\Delta=0.1$, the system bifurcated from the IS to SS via a first-order transition (IS$\rightarrow$SS). When it is large, for example, $\Delta=0.5$, the system transfers from the IS to SS via BS (IS$\rightarrow$BS$\rightarrow$SS). In this case, similar to Fig. \ref{fig1}(b), there are two stages within the regime of BS. In Fig. \ref{fig7}, a typical BS is shown at $K=1.80$, which belongs to the second stage of BS regime.
It is seen that the dynamical features are similar to that of the BS in Fig. \ref{fig3}.

\begin{figure}[htbp]
\begin{center}
\includegraphics[width=0.5\textwidth]{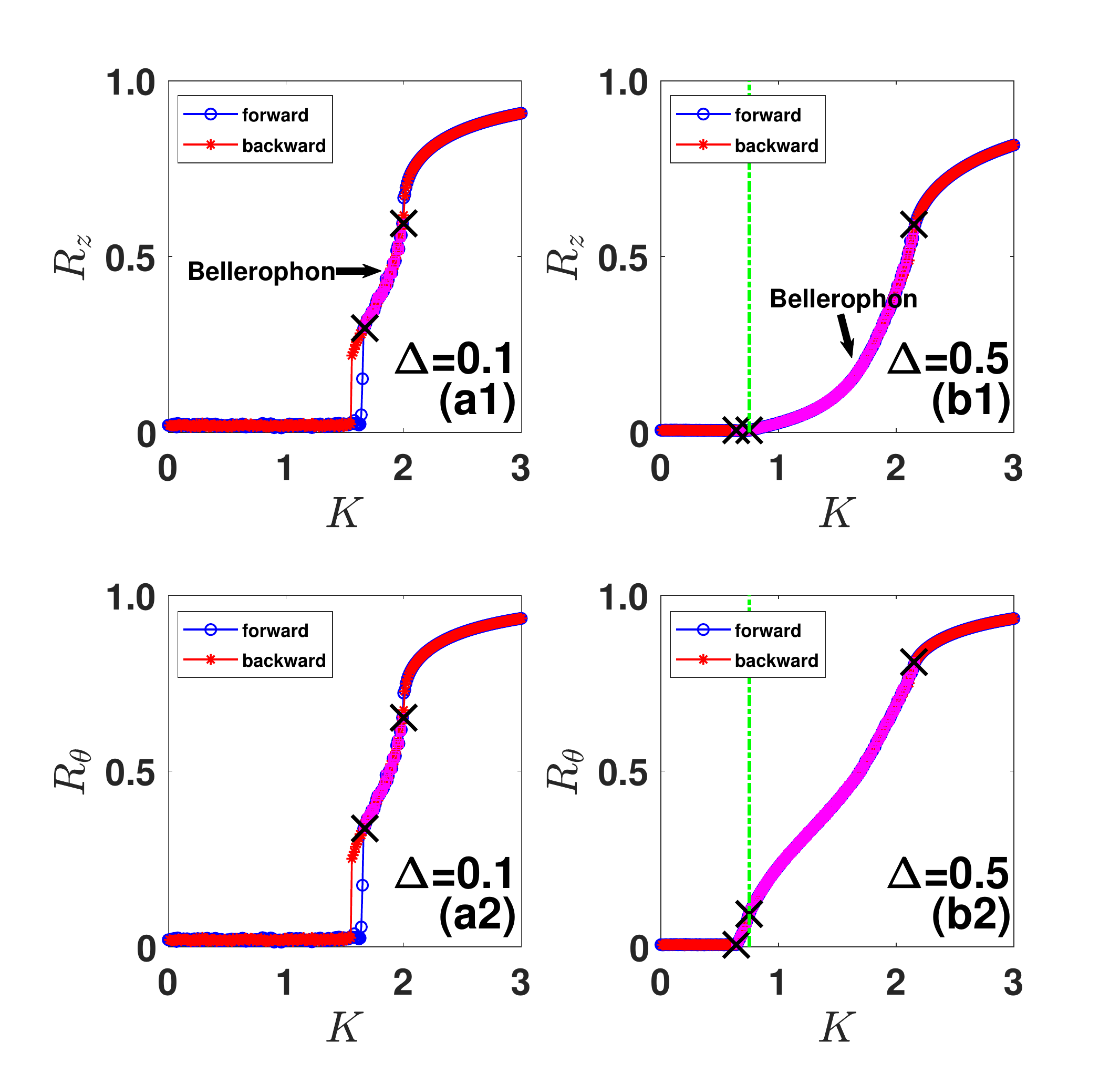}
\caption{(Color online).
Typical synchronization transition paths in coupled SL oscillators (Eq. \ref{eq:model}) with the uniform frequency distribution. For $\Delta=0.1$, the transitions towards synchronization is hybrid. The three transition points are
$K=1.57$, $1.65$, and $2.02$, respectively.
For $\Delta=0.5$, the transition is continuous, and the three transition points are $K=0.64$, $0.75$, and $2.16$ (marked by the crosses), respectively
The regime of BS is denoted by the magenta color, and the green line marks the transition point within BS regime.
}\label{fig8}
\end{center}
\end{figure}

Finally, we briefly present the results for the uniform frequency distribution. When $\Delta$ is small, for example, $\Delta=0.1$ as shown in Fig.~\ref{fig8}(a), the system first bifurcates from the IS  to BS and then to SS, i.e., the bifurcation path is IS$\rightarrow$BS$\rightarrow$SS as the coupling strength increases. It is found that IS$\rightarrow$BS is  first-order while BS$\rightarrow$SS is continuous. The same situation happens as the coupling strength decreases in the backward transition and thus form a hysteresis loop.
When $\Delta$ is large, for example, $\Delta=0.5$ as shown in Fig.~\ref{fig8}(b), the system's bifurcation path is IS$\rightarrow$BS$\rightarrow$SS as the coupling strength increases, and both transitions are continuous. Similar to the previous two distributions, two stages are observed within the BS regime. The above results suggest that the transition within the BS regime is due to the inherent amplitude dynamics of SL oscillators, despite of specific frequency distributions.

\section{Conclusion}

Bellerophon state is a higher order coherent state, in which oscillators form quantized coherent clusters, and in each coherent cluster the oscillators' instantaneous frequencies are not locked, but their average frequencies are locked instead. Previously, it has been found that such states are  generic in coupled phase oscillators.
In this work, we investigated the synchronization in coupled SL oscillators, and found that
Bellerophon state also exists in such system with amplitude dynamics. Depending on the parameter characterizing the width of unimodal frequency distribution, both first-order and second-order transitions have been observed. Typically, the system bifurcates from IS to SS via BS when the frequency distribution is significantly wide. Interestingly, we revealed that there is a novel transition with the regime of BS, i.e., from chaotic phase synchronization to periodic phase synchronization. The present work suggests that there might be more higher order collective behaviors in coupled oscillator systems when the amplitude dynamics is involved.

\section*{Acknowledgments}
This work is partially supported by the National Natural
Science Foundation of China (Grants No. 11875132, No. 11835003, and No. 11872182),  and the Natural Science Foundation of Shanghai
(Grants No. 18ZR1411800 and No. 17ZR1444800).

\end{document}